\documentclass[twocolumn,reprint,amsmath,amssymb,aps,prm,superscriptaddress,nobibnotes]{revtex4-1}
\usepackage{graphicx}

\begin{document}

\title{Giant Rashba spin splitting in strained KTaO$_3$ ultrathin films for circular photogalvanic currents}

\author{Ning Wu}
\affiliation{Beijing National Laboratory for Condensed Matter Physics, Institute of Physics, Chinese Academy of Sciences, Beijing 100190, China}
\affiliation{School of Physical Sciences, University of Chinese Academy of Sciences, Beijing 100190, China}
\author{Xue-Jing Zhang}
\email[]{Present affiliation: Institute for Advanced Simulation, Forschungszentrum J\"ulich, D-52425 J\"ulich, Germany}
\affiliation{Beijing National Laboratory for Condensed Matter Physics, Institute of Physics, Chinese Academy of Sciences, Beijing 100190, China}
\affiliation{School of Physical Sciences, University of Chinese Academy of Sciences, Beijing 100190, China}
\author{Bang-Gui Liu}
\email[]{Email: bgliu@iphy.ac.cn}
\affiliation{Beijing National Laboratory for Condensed Matter Physics, Institute of Physics, Chinese Academy of Sciences, Beijing 100190, China}
\affiliation{School of Physical Sciences, University of Chinese Academy of Sciences, Beijing 100190, China}

\date{\today}

\begin{abstract}
Strong Rashba effects at surfaces and interfaces have attracted great attention for basic scientific exploration and practical applications.
Here, the first-principles investigation shows that giant and tunable Rashba effects can be achieved in KTaO$_3$ (KTO) ultrathin films by applying biaxial stress. When increasing the in-plane compressive strain nearly to -5\%, the Rashba spin splitting energy reaches $E_{R}=140$ meV, approximately corresponding to the Rashba coupling constant $\alpha_{R}=1.3$ eV {\AA}. We investigate its strain-dependent crystal structures, energy bands, and related properties, and thereby elucidate the mechanism for the giant Rashba effects. Furthermore, we show that giant Rashba spin splitting can be kept in the presence of SrTiO$_3$ capping layer and/or Si substrate, and strong circular photogalvanic effect can be achieved to generate spin-polarized currents in the KTO thin films or related heterostructures, which are promising for future spintronic and optoelectronic applications.
\end{abstract}
\maketitle
\maketitle

\section{Introduction}

The Rashba spin-orbit interaction\cite{rashba1960ei,bychkov1984oscillatory,bychkov1984properties}
due to the broken inversion symmetry and the atomic spin-orbit coupling (SOC) can result in the momentum-dependent spin splitting of the electron states, which can be used as an effective way for spin manipulation.
Rashba effect plays key roles in quantum wells\cite{nestoklon2008electric}, two dimensional (2D) electron gases (2DEG)\cite{lesne2016highly}, and thin films based on traditional III-V semiconductors\cite{PhysRevLett.78.1335}.
After intensive investigations, one can tailor the Rashba coupling by electric field and strain, and design artificial microstructures for wide applications. External electric field can be used to modulate the magnitude of Rashba spin splitting in LaAlO$_3$/SrTiO$_3$ (LAO/STO) interface\cite{PhysRevLett.104.126803} and InSe multilayer \cite{premasiri2018tuning}.
The Rashba spin splitting  can be effectively tuned by varying the interlayer distance in graphene/As-I van der Waals heterostructure\cite{yu2018tunable} and adjusting the halogen doping concentration in doped PtSe$_2$ monolayer\cite{absor2018strong}. It is very interesting to manipulate the Rashba spin-orbit coupling by applying strain, as were done in 2D LaOBiS$_2$\cite{liu2013tunable}, binary alloyed hexagonal nanosheets\cite{zhu2018huge}, 2D heterostructures\cite{zhang2018rashba}, and BiSb monolayer\cite{singh2017giant}.

Recently, a 2DEG was observed at KTaO$_3$ (KTO) (100) surface\cite{king2012subband,santander2012orbital}, but the Rashba spin splitting of the 2DEG was not resolved from the angle-resolved photoemission (ARPES) spectrum\cite{king2012subband}. For another 2DEG at an amorphous-LAO/KTO interface, an experimental analysis of the weak anti-localization effect resulted in a Rashba coupling constant 0.1 eV \AA{}, and a 50-fold enhanced Hall mobility of charge carriers with Rashba SOC was achieved\cite{zhang2019hui}. Surprisingly, hysteretic magnetoresistance up to 25 K and anomalous Hall effect up to 70 K were observed at an EuO/KTO interface\cite{zhang2018high}. Theoretically, the magnitude of Rashba spin splitting in KTO surface was studied by applying external electric fields in a symmetrical slab model\cite{PhysRevLett.112.086802}. It is a challenge, however, to avoid the possible background noise or short circuit in the measurement especially for nanospintronic devices because such external electric fields usually need a power supply\cite{zhu2018piezotronic}. It is interesting to investigate the effects of in-plane strain fields on the strength of Rashba spin splitting for KTO surfaces. Actually, strain (stress) is a wonderful approach to manipulate the crystal structures of KTO and thus control their electronic structures and functional properties. Recent studies have demonstrated that the strain can affect the formation and migration of oxygen vacancies in KTO\cite{xi2017strain} and drive electron-hole interchanging of the two opposite surface 2D carrier gases in KTO ultrathin film\cite{Zhang2018Strain}.

Here, through first-principles calculations and further analyses, we investigate the in-plane strain dependencies of the structural features, intrinsic electrostatic potentials, band edges, carrier concentrations, carrier effective masses, and Rashba parameters of the KTO slabs. We show that the Rashba spin splitting of ultrathin KTO films can be controlled by applying biaxial stress and thus giant Rashba-like spin splitting can be obtained by applying compressive biaxial stress. In addition, we explore circular photogalvanic currents due to the giant Rashba-like spin splitting in the ultrathin KTO films. More detailed results will be presented in the following.

\begin{figure*}
\centering
\includegraphics[width=0.9\textwidth]{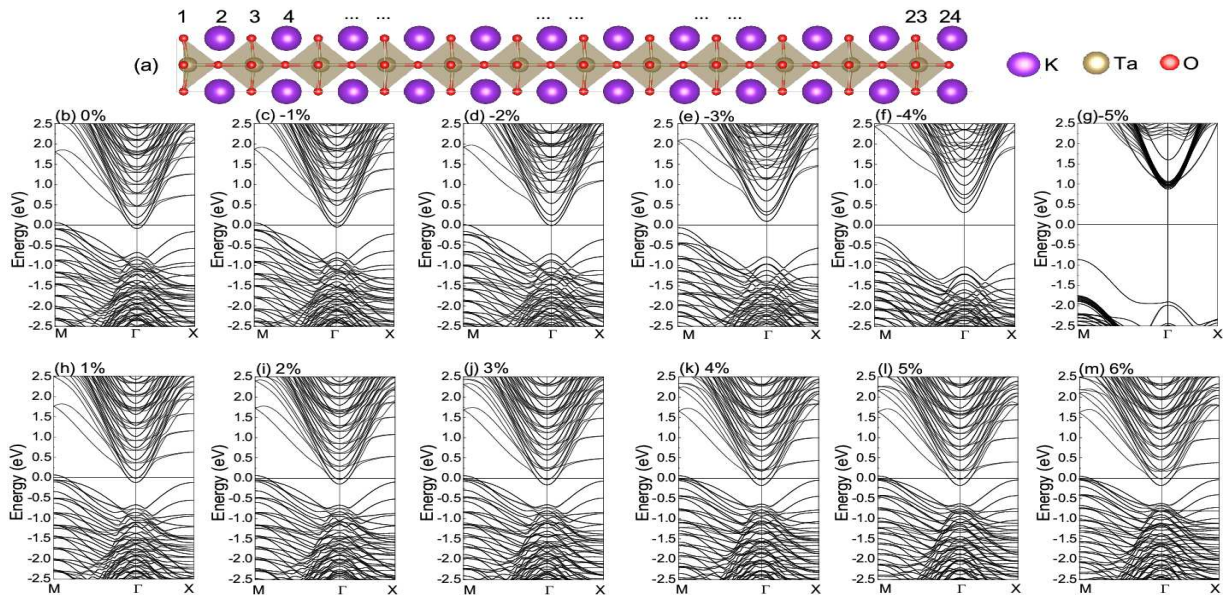}
\caption{\label{Fig1}
(a) Side view of the optimized atomic structure of the KTO ultrathin film (slab) at $\varepsilon_{s}$ = 0\%. (b-m) Band structures of the KTO ultrathin film at different strain values: $\varepsilon_{s}$ = 0\%, -1\%, -2\%, -3\%, -4\%, -5\%, 1\%, 2\%, 3\%, 4\%, 5\%, and 6\%. The K, Ta, and O atoms are shown by the purple, yellow, and red balls, respectively. The Fermi level $E_{f}$ is at the zero energy. }
\end{figure*}

\section{Computation method }

Our first-principles calculation is performed using the projector-augmented wave method within the density-functional theory\cite{hohenberg1964inhomogeneous,kohn1965self}, as implemented in the Vienna Ab initio Simulation Package (VASP)\cite{kresse1999ultrasoft,blochl1994projector}. To describe the exchange-correlation energy, we used the general gradient approximation (GGA) with the Perdew-Burke-Ernkzerhof for solids (PBEsol) parametrization\cite{perdew1996generalized,perdew2008restoring}.
The on-site Coulomb interaction in 5d states of transition-metal ions is corrected by the DFT+$U$ (where $U$ is the Hubbard energy) method\cite{anisimov1997first}. The effective value $U_{eff} = 3$ eV is employed for Ta 5d states in this work, as it is well established that such a value is appropriate to describe these strongly-correlated states\cite{Zhang2018Strain}.
An Monkhorst-Pack k-point grid of $4\times4\times1$ is used for reciprocal space integration, and the plane wave energy cutoff is set to 500 eV. Our convergence standard requires that the Hellmann-Feynman force on each atom is less than 0.01 eV/{\AA} and the absolute total energy difference between two successive consistent loops is smaller than $1\times10^{-5}$ eV. A fully converged electronic structure is used for further calculation including SOC. A 20 {\AA} thick vacuum layer is used in the KTO-slab geometry. Additional calculations with vacuum layer of 30\AA{} and dipole corrections\cite{DFTPolar} are made for confirmation.
When a biaxial stress is applied, the in-plane strain is defined as $\varepsilon_{s}$=($a-a_{0}$)/$a_{0}$ $\times100\%$, where $a_{0}$ is the experimental lattice constant of bulk KTO without strain ($a_{0} = 3.989$ {\AA}\cite{peng2014measurement}) and $a$ is the in-plane lattice constant of strained KTO slab. Given an in-plane strain value, the out-of-plane lattice and all the internal atomic positions are allowed to relax sufficiently during optimization.

\section{RESULTS AND DISCUSSION }

\subsection{KTO slab under biaxial stress}

We construct a KTO slab model to describe the KTO ultrathin film under different biaxial stresses. The slab consists of $m=12$ KTO unit cells along the vertical [001] axis. Fig. \ref{Fig1}(a) shows the optimized structure of the KTO slab at the in-plane strain $\varepsilon_{s}$ = 0\% (zero stress). We study the strained KTO slabs with the in-plane strain $\varepsilon_{s}$ ranging from -5\% (compressive) to +8\% (tensile). With a given in-plane strain, the system is fully optimized, with the out-of-plane strain being determined by requiring that the out-of-plane stress is zero, and thus we can determine the in-plane stress. Actually, this is a system with biaxial stress. With the condition that the out-of-plane stress is zero, however, the in-plane stress is determined by the in-plane strain. Therefore, for convenience, we shall use the in-plane strain $\varepsilon_{s}$ to characterize the strained slabs in the following. It is confirmed that the dipole correction has little effect in these results.
In Fig. \ref{Fig1}(b-m), we plot the representative electronic band structures along M ($\pi$,$\pi$)$\rightarrow$ $\Gamma$ (0,0) $\rightarrow$ X (0,$\pi$) of the optimized KTO slabs for $\varepsilon_{s}$ = 0\%, -1\%, -2\%, -3\%, -4\%, -5\%, 1\%, 2\%, 3\%, 4\%, 5\%, and 6\%, respectively. Here, SOC is taken into account, and the k vector is in units of 1/$a$, where $a$ is the calculated lattice constant of the strained KTO slab.
From Fig. \ref{Fig1}(g,f,e), it is clear that the KTO slab is insulating for $\varepsilon_{s}$ = -5\%, -4\%, and -3\%. There is a critical strain $\varepsilon_{s}$ = -2\%, as shown in Fig. \ref{Fig1}(d). When $\varepsilon_{s}$ becomes larger than -2\%, the KTO slab is metallic. Therefore, there is a strain-driven insulator-to-metal phase transition at $\varepsilon_{s}$ = -2\%.
For the metallic state, there are electron carriers near the $\Gamma$ point and hole carriers near the M point, and they form a 2DEG at the TaO$_2$-terminated surface and a 2D hole gas (2DHG) at the KO-terminated surface. It should be pointed out that the electron concentration in the 2DEG is equivalent to the hole concentration in the 2DHG\cite{Zhang2018Strain}.

\begin{figure}
\centering
\includegraphics[width=0.45\textwidth]{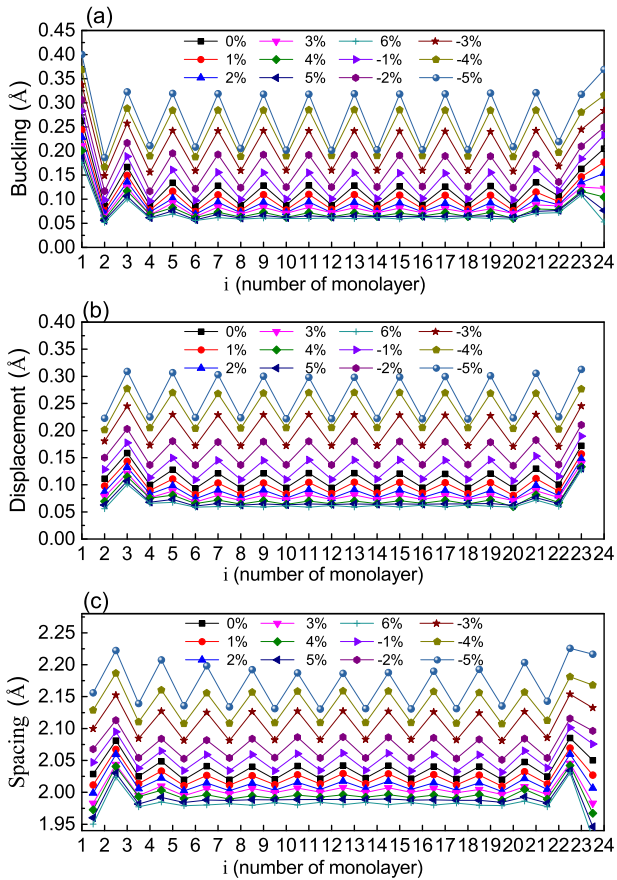}
\caption{\label{Fig2}
The monolayer-resolved intra-monolayer bucklings (a), displacements of cations (Ta, K) with respect to the nearest O anions (b), inter-monolayer spacings (c) of the KTO slab at the different $\varepsilon_{s}$ values.
}
\end{figure}

\begin{figure}
\centering
\includegraphics[width=0.36\textwidth]{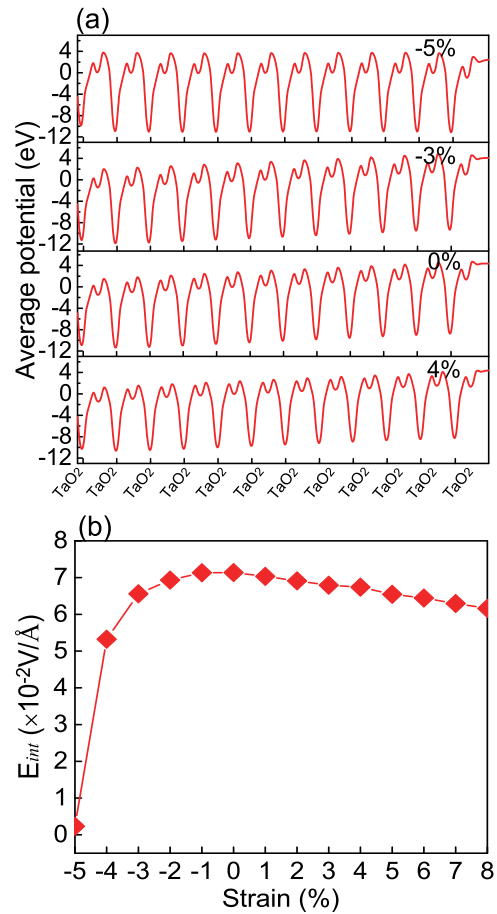}
\caption{\label{Fig3}
(a) The monolayer-resolved plane-averaged electrostatic potentials of the KTO slab for strain $\varepsilon_{s}$ = -5\%, -3\%, 0\%, and 4\%, respectively. (b) The internal electric field ($E_{int}$) as a function of the strain $\varepsilon_{s}$.}
\end{figure}

To show the stress-driven structural features, we present in Fig. 2 the monolayer-resolved intra-monolayer ionic bucklings (defined as the maximal cation-anion out-of-plane difference within the monolayer), out-of-plane cation displacements with respect to the centers of the nearest O anions, and inter-monolayer spacings of the slab under the in-plane strain values between -5\% and 6\%. It is clear that the surfaces make big changes with respect to the internal region in the three aspects, and in the internal region the three aspects are made nearly independent of monolayers at strong tensile strains. It is interesting that all the three values monotonically decrease with tensile strain, but increase with compressive strain. When the compressive in-plane strain becomes strong, the bucklings of the surface monolayers are substantially enhanced, and the displacements become nearly the same value $d_1$ for all the KO monolayers or $d_2$ for all the TaO$_2$ monolayers, with $d_2>d_1$. By combining the buckling and spacing values, it is visible that the two surface single-unit-cells are a little separated from the main body.

Furthermore, we plot in Fig. 3(a) the plane-averaged electrostatic potentials for $\varepsilon_{s}$ = -5\%, -3\%, 0, and 4\% as representative strain values. It is clear that the maximal (or minimal) value increases from the left to right hand side in the cases of $\varepsilon_{s}$ = -3\%, 0, and 4\%, but remains the same for $\varepsilon_{s}$ = -5\%. The internal electric field $E_{int}$ can be estimated from the slope of the plane-averaged electrostatic potential shown in Fig. 3(a)\cite{choe2018band}. The calculated results as a function of $\varepsilon_{s}$ are presented in Fig. 3(b). The internal electric field at the unstrained KTO slab is 7.1$\times 10^{-2}$ V/{\AA}, comparable with a previous study\cite{stengel2009berry}. It is clear that $E_{int}$ slowly decreases with tensile strain, and accelerates with compressive strain, nearly reaching zero at $\varepsilon_{s}$ = -5\%.
It is expected that  the out-of-plane cation displacements with respect to the neighboring O atoms counteract the out-of-plane polarity of the KTO slab, which originates from the oppositely charged (TaO$_{2}$)+ and (KO)- monolayers.
When the compressive strain reaches $\varepsilon_{s}$ = -5\%, corresponding to the in-plane lattice constant 3.789 {\AA}, the potential slope is almost diminished by the increasing polarization due to displacements. The large changes in $E_{int}$ caused by strong compressive strains will change the energy bands.

\subsection{Energy band parameters}

\begin{figure}
\centering
\includegraphics[width=0.48\textwidth]{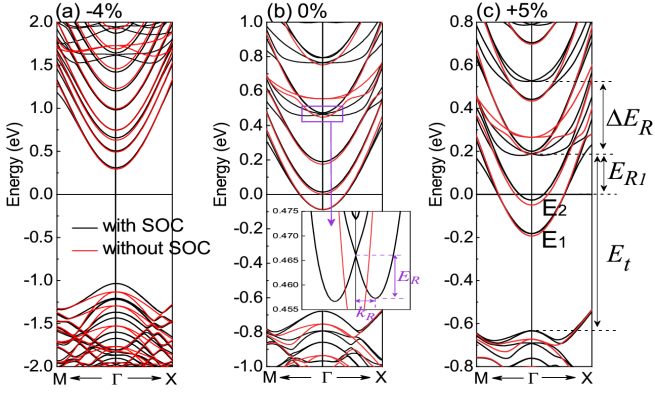}
\caption{\label{Fig4}
The band structures around $\Gamma$ point of the KTO slab with (black) and without (red) SOC for $\varepsilon_{s}$ =-4\%, 0\%, and 5\% are magnified in (a), (b) and (c), respectively. The inset of (b) shows the definition of $E_{R}$ and $k_{R}$ used for estimating the Rashba parameters. The lowest and second lowest $d_{xy}$ bands are labeled by E$_{1}$ and E$_{2}$.}
\end{figure}

\begin{figure}
\centering
\includegraphics[width=0.4\textwidth]{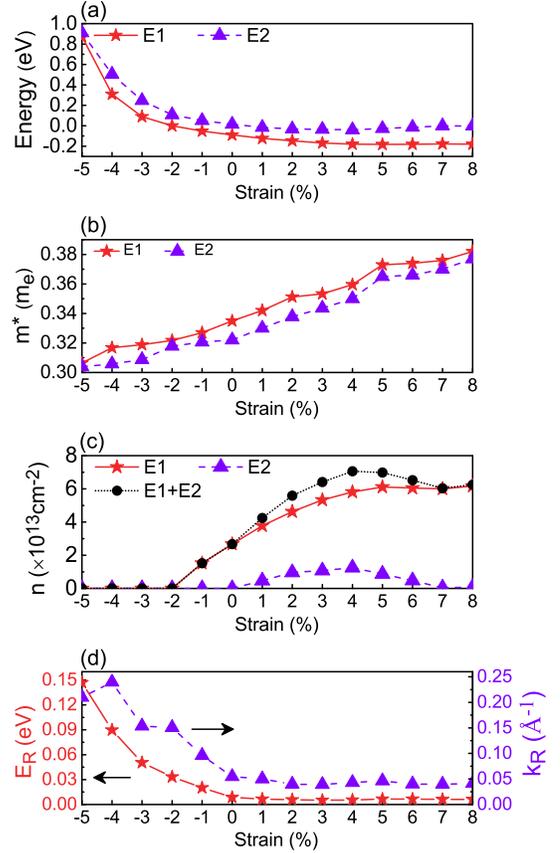}
\caption{\label{Fig5}
The strain $\varepsilon_{s}$ dependencies of the energy positions of the E$_{1}$ and E$_{2}$ bands (a), the corresponding electron effective masses $m^\star$  (b), the 2DEG concentration $n$ (c), and the Rashba spin splitting energy $E_R$ and k-vector offset $k_R$ (d). The E$_{1}$ and E$_{2}$ bands are defined  in Fig. 2.}
\end{figure}

To elucidate the band edges and electron concentrations for the strained KTO slab, we present in Fig. 4 (red lines) the magnified electron band structures without SOC of the KTO slab near the $\Gamma$ point for $\varepsilon_{s}$ = -4\%, 0\%, and 5\%. In the absence of SOC, the quantum confinement reduces the initial cubic symmetry of the Ta $t_{2g}$ orbitals in the bulk perovskite. The triple degeneracy (excluding spin) of the $t_{2g}$ bands at the $\Gamma$ point is lifted, splitting the $d_{xy}$ from $d_{xz}$/$d_{yz}$.
When the inversion symmetry breaking is taken into account, the Ta atomic SOC further splits the $d_{xz}$/$d_{yz}$ bands into the upper part and the lower one, except for the time-reversal invariant momenta: $\Gamma$, X, and M.
For the $\varepsilon_{s}$ = 0\% case shown in Fig. 4(b), only the lowest $d_{xy}$ band is partially occupied and the band minimum (at the zone center) lies 0.089 eV below $E_{f}$. The corresponding electron concentration of the 2DEG is $2.67\times 10^{13} $cm$^{-2}$, which is an order of magnitude smaller than $2\times$ 10$^{14}$cm$^{-2}$ of the 2DEG formed at an experimental KTO surface from the ARPES measurements\cite{king2012subband}.
This difference can be interpreted by the low formation energy for the oxygen vacancies at the KTO surface\cite{king2012subband,wadehra2017electronic}, which allows much more electrons in the 2DEG.
For the $\varepsilon_{s}$ = 5\% case shown in Fig. 4(c), the minima of the occupied lowest and second lowest $d_{xy}$ states (at the $\Gamma$ point) are 0.182 and 0.027 eV below $E_{f}$, in which the summed electron concentration is 6.98 $\times 10^{13}$cm$^{-2}$, larger than that in the unstrained system.
For the $\varepsilon_{s}$ = -4\% case shown in Fig. 4(a), the lowest $d_{xy}$ band lies 0.31 eV above the $E_{f}$, which means that there are no carriers, in contrast with those of the $\varepsilon_{s}$ = 0\% and 5\% cases.

It is obvious that there are some Rashba-like spin splitting in the conduction bands of the KTO slab for  the $\varepsilon_{s}$ between -5\% and 8\%. To further investigate the Rashba effects, we also present in Fig. 4 the magnified electron band structures with SOC of the KTO slab near the $\Gamma$ point (black lines) for $\varepsilon_{s}$ = -4\%, 0\%, and 5\%.

To better describe the properties of the lowest and second lowest $d_{xy}$ conduction bands at the $\Gamma$ point respectively defined by $E_{1}$ and $E_{2}$ in Fig. 4(c), the band edge positions, electron effective masses, and  2DEG concentrations of the $E_{1}$ and $E_{2}$ bands as  functions of $\varepsilon_{s}$ are calculated and shown in Fig. 5. In Fig. 5(a), as $\varepsilon_{s}$ changing from -5\% to 4\%, the band edge positions of E$_{1}$ and E$_{2}$ decrease rapidly for large compressive strain, but they change slowly for tensile strain. As $\varepsilon_{s}$ varying from 5\% to 8\%, the band edge positions of $E_{1}$ and $E_{2}$ are almost unchanged. It should be noted that the bottom of the conduction band is less affected by the tensile strain, while it is significantly changed by the compressive strain.
In detail, the band edge of the $E_{1}$ band is below $E_{f}$ for $\varepsilon_{s} \geq $-1\%, and that of $E_{2}$ becomes below $E_{f}$ for $\varepsilon_{s} \geq$ 1\%.

In Fig. 5(b), the effective mass ($m^*$) is evaluated from a second-order fit of the band energies using $m^* = \frac{\hbar^2}{d^2E(k)/dk^2}$\cite{xia2018universality}.
Remarkably, the values of $m^*_1$ for $E_{1}$ and $m^*_2$ for $E_{2}$ in the unstrained KTO system are 0.35 and 0.32 $m_{e}$ ($m_{e}$ is the mass of the free electron), respectively, which are both in excellent agreement with 0.30 $m_{e}$ for the KTO surface 2DEG measured by ARPES\cite{king2012subband}.
When the strain changing from $\varepsilon_{s}$ = -5\% to 8\%, $m^*_1$ and $m^*_2$ increase, with $m^*_2$ being always smaller than $m^*_1$.
Under $\varepsilon_{s}$ = 8\%, $m^*_1$ and $m^*_2$ reach the maximum values 0.38 and 0.37 $m_{e}$, respectively, which are still smaller than $\sim$ 0.5 and 0.6 m$_{e}$ recently determined for a surface 2DEG on STO\cite{meevasana2011creation}. This suggests that developing high-mobility oxide electronics by KTO is better than by STO.

Fig. 5(c) shows the relationship between $\varepsilon_{s}$ and the carrier concentrations.
For $\varepsilon_{s} \leq -2\%$, the $E_{1}$ and $E_{2}$ bands are empty and the KTO film is insulating, which is consistent with the critical strain of insulator-metal transition shown in Fig. 1.
The carrier concentrations $n_1$ and $n_2$ for the $E_{1}$ and $E_{2}$ bands have the maximum values at the $\varepsilon_{s} = 5\%$ and $4\%$, respectively, and the total 2DEG concentration $n$ of the E$_{1}$ + E$_{2}$ bands reaches the maximum values of 7.06 $\times 10^{13}$cm$^{-2}$ at $\varepsilon_{s}$ = 4\%.
This indicates that the conductivity of the 2DEG formed at the surface can be effectively modulated by the in-plane strain.

In addition, we summarize in Table I the energy differences ($E_{R1}$) between the first Rashba band minimum and the Fermi level [and valence band edge ($E_t$)] and the spin-orbit splitting energy $\Delta E_{R1}$ between the lowest and the second lowest Rashba doublets, as defined in Fig. 4.

{\renewcommand\arraystretch{1.5}
\begin{table}
\caption{\label{Table1} The three band parameters [$E_{R1}$, $E_{t}$, and $\Delta E_{R1}$ (eV)] and the two Rashba parameters [$E_{R}$ (meV) and $k_{R}$ (\AA{}$^{-1}$)] of the KTO ultrathin film at different strains. }
\begin{ruledtabular}
\begin{tabular}{cccccc}
strain & $E_{R1}$ & $E_{t}$ & $\Delta E_{R1}$ & $E_{R}$ & $k_{R}$\\
\hline
$\varepsilon_{s}=5\%$ & 0.188 & 0.822 & 0.339 & 6.6 & 0.046  \\
$\varepsilon_{s}=0\%$ & 0.466 & 1.144 & 0.298 & 8.6 & 0.054  \\
$\varepsilon_{s}=-1\%$ & 0.603 & 1.282 & 0.309 & 20. & 0.096  \\
$\varepsilon_{s}=-2\%$ & 0.810 & 1.516 & 0.309 & 33. & 0.151  \\
$\varepsilon_{s}=-3\%$ & 1.161 & 1.947 & 0.314 & 51. & 0.154 \\
$\varepsilon_{s}=-4\%$ & 1.618 & 2.652 & 0.325 & 90. & 0.24  \\
$\varepsilon_{s}=-5\%$ & 2.228 & 4.136 & 0.093 & 140 & 0.21
\end{tabular}
\end{ruledtabular}
\end{table}
}

\subsection{Rashba spin splitting}

Since the KTO slab obeys the $C_{4v}$ point group symmetry, the symmetry-allowed linear spin-momentum coupling can be expressed as\cite{bychkov1984properties} $H_{R}=\alpha_{R}(k_{x}\sigma_{y}-k_{y}\sigma_{x})$.
According to the linear Rashba model, the dispersion due to the Rashba spin splitting can be described by
\begin{eqnarray}
E=\frac{\hbar^2}{2m^*}(k\pm k_{R})^2-E_{R}, \label{eq:one}
\end{eqnarray}
where $k$ is the magnitude of the electron wave vector, $m^*$ is the electron effective mass, $E_{R}=\hbar^2k_R^2/2m^*$ is the Rashba spin splitting energy, and $k_{R}$ is the momentum offset. The in-plane spin polarizations of the "+" and "-" eigenstates are oppositely aligned and normal to the electron wave vector.
In the isotropic case, the Rashba coupling constant can be estimated by $\alpha_{R}$ = $2E_{R}/k_{R}$, and $\alpha_{R}$ depends on the strength of SOC and inversion asymmetry\cite{zhong2015giant}.

For the KTO slab, the lowest Rashba spin split bands near the $\Gamma$ point are similar to those defined by Eq. (1), and we present $E_{R}$ and $k_{R}$ in Fig. 5(d) for different in-plane strains.
The calculated values of $E_{R}$ and $k_{R}$  are summarized in Table I for $\varepsilon_{s}$ = 5\%, 0\%, -1\%, -2\%, -3\%, -4\%, and -5\%.
In Fig. 5(d), $E_{R}$ and $k_{R}$ increase drastically with the compressive strain increasing, but they are both almost unchanged for increasing tensile strain. In Table I, noticeably, $E_{R}$ and $k_{R}$ are 140 (90) meV and 0.21 (0.24) {\AA}$^{-1}$ for the KTO slab at $\varepsilon_{s}= -5$\% (-4\%). It is clear that compressive in-plane strain can enhance the Rashba spin splitting energy $E_{R}$. Because $E_{int}$ is near zero at $\varepsilon_{s} = -5.0$\%, $E_{int}$ becomes negative when $\varepsilon_{s}< -5.0$\%, as shown in Fig. 3, and consequently the conduction bands are reconstructed, which leads to smaller $E_{R}$ or substantial deformation of the Rashba bands. Actually, this means that the maximal $E_{R}$ is reached at $\varepsilon_{s}$ = -5\%.

For comparison, we summarize the $E_{R}$, $k_{R}$, and $\alpha_R$ values of some typical Rashba systems in Table II.
For brevity, we can take $E_{R}$ as the key parameter to characterize such Rashba systems. $E_{R}$ can reach 100 meV for BiTeI van der Waals bulk \cite{ishizaka2011giant}, or 190 meV for
$\alpha$-GeTe(111) film \cite{krempasky2016disentangling}. In contrast, for perovskite oxides, the previous maximal $E_{R}$ is 15 meV for KTaO$_3$/BaHfO$_3$ interface \cite{kim2016strongly}. It is clear that our strategy is very efficient to promote the Rashba spin splitting energy in perovskite oxides because our maximal $E_{R}$ value reaches 140 meV at $\varepsilon_{s}$ = -5\%.

{\renewcommand\arraystretch{1.5}
\begin{table}
\caption{\label{Table2} Rashba splitting energy $E_{R}$ (meV), k-vector offset $k_{R}$ ({\AA}$^{-1}$), and Rashba coupling constant $\alpha_{R}$ (eV~{\AA}) of typical 2D materials (monolayer and van der Waals multilayers), sp semiconductors, and perovskite oxides.}
\begin{ruledtabular}
\begin{tabular}{cccc}
system & $E_{R}$ & $k_{R}$ & $\alpha_{R}$\\
\hline
GaSe/MoSe$_2$ van der Waals HS \cite{zhang2018rashba} & 31  & 0.13 & 0.49  \\
BiTeI monolayer ($\varepsilon_{s}=6\%$) \cite{ma2014emergence} & 55.7  & 0.054 & 2.05  \\
BiTeI van der Waals bulk \cite{ishizaka2011giant} & 100  & 0.052 & 3.8  \\ \hline
InAlAs/InGaAs interface \cite{PhysRevLett.78.1335} & $<$1.0 & 0.028 & 0.07  \\
GeTe(111)/InP(111) interface \cite{meng2017ferroelectric} & 5.403  & 0.010 & 1.08  \\
$\alpha$-GeTe(111) film \cite{krempasky2016disentangling} & 190 & 0.13 & 4.2  \\ \hline
BiAlO$_3$ bulk crystal \cite{da2016rashba} & 7.34  & 0.038 & 0.39  \\
LaAlO$_3$/SrTiO$_3$ interface \cite{PhysRevLett.104.126803} & $<$5.0 &  & 0.01$\sim$0.05  \\
KTaO$_3$/BaHfO$_3$ interface \cite{kim2016strongly}   & 15 & & 0.3  \\
KTaO$_3$ film ($\varepsilon_{s}=-5\%$)  & 140  &  0.21 & 1.3 \\
\end{tabular}
\end{ruledtabular}
\end{table}
}

\subsection{Heterostructures and photogalvanic effect}

For real applications, the KTO surfaces could be capped with STO overlayers, and/or KTO films should be grown on good substrates. For this purpose, we study KTO/STO superlattices for the two STO/KTO interfaces. For (STO)$_{4}$/(KTO)$_{12}$ superlattice with $a_{\parallel} = a_{STO}$ and $a_{KTO}$, the bottom of the conduction bands and the lowest of the Rashba spin splitting band are respectively arising from the Ta $d_{xy}$ and Ta $d_{xz}$/$d_{yz}$ states for the TaO$_{2}$ monolayer at the SrO/TaO$_{2}$ interface, and top of valence bands are derived from the O p states for the TiO$_{2}$ monolayer at the TiO$_{2}$/KO interface. This indicates that the band gap in all the superlattices is formed between two spatially separated bands, and the carriers at the valence band can not be excited by the circularly polarized light to the conduction band. This implies that capping (making the interface) can diminish the carriers in the bare surface. The metallic interfaces can be obtained by generating carriers through
experimentally applying gate voltage at room temperature\cite{thiel2006tunable}. We also study STO/KTO/Si trilayer to simulate capped KTO films on Si substrate.

Furthermore, the STO capping affects the Rashba spin splitting in the bare TaO$_2$-terminated surfaces of KTO films. For the in-plane strains of -3\%, -4\%, and -5\%, our calculations show that the maximal Rashba spin splitting energies are 25, 95, and 190 meV, respectively, and the corresponding k vector offsets are 0.12, 0.19, and 0.24 {\AA}$^{-1}$. For STO/KTO/Si trilayer, the lattice mismatch is 3.6\% for STO/KTO on Si substrate, and then $\varepsilon_{s}$ is $-3.6\%$ and the Rashba spin splitting energy is 51 meV, with the k vector offset being 0.16 {\AA}$^{-1}$. It is clear that the Rashba effects are still very strong after capping layers and/or substrates are added.

\begin{figure}
\centering
\includegraphics[width=0.4\textwidth]{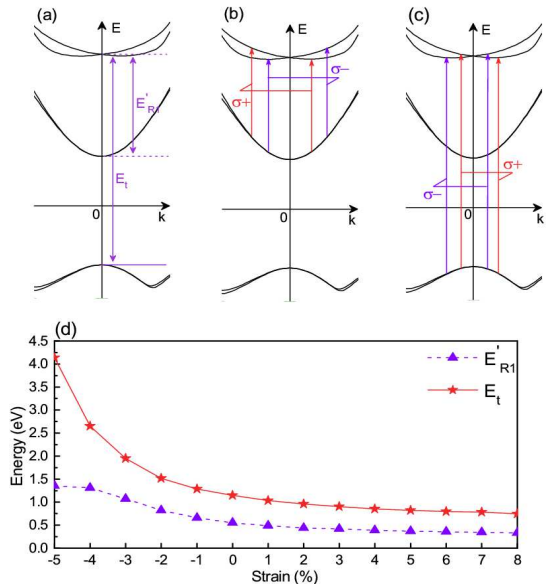}
\caption{\label{Fig6}
Schematic for CPGE: photon energy for transition from the valence band edge ($E_t$) or from the conduction band edge ($E^\prime_{R1}$) to the Rashba bands (a); the two permitted transitions (only two $k_x$ values, $k^{\pm}_{x}$, for a given $k_y$) from the conduction band edge through circularly polarized light with $\sigma_+$ (red) or $\sigma_-$ (purple) (b), and the similar transitions from the valence band edge (c); and the strain dependence of the needed photon energies $E^\prime_{R1}$ and $E_t$ (d).}
\end{figure}

Because of the giant Rashba spin splitting, the KTO ultrathin films can be used for achieving circular photogavalnic effect (CPGE) to generate spin-polarized photocurrents\cite{ganichev2001conversion,moayed2017towards}. For the right-handed (left-handed) circularly polarized light, its photon has the angular momentum of +1 (-1), labeled by $\sigma_{+}$ ($\sigma_{-}$), and the selection rule for necessary transitions is that the allowed z-component change of the total angular momentum is $\Delta J_m=$+1 (-1). The valence band edge, originating from the $J=3/2$ states, has $J_m=\pm 3/2$, and the conduction band edge, from the $d_{xy}$ states of $l_m=2$ and $-2$, consists of $J_m=\pm 3/2$ and $\pm 5/2$ states. Because the $d_{yz}$/$d_{xz}$ bands have $l_m=1$ and $-1$, the Rashba split bands consist of $J_m=\pm 3/2$ and $\pm 1/2$ states. Therefore, for achieving the CPGE, the electrons can transit from the valence band top (with photon energy $E_t$) or the filled conduction band edge ($E^\prime_{R1}$) to the $d_{yz}$/$d_{xz}$-based bands with giant Rashba spin plitting as the final states, as shown in Fig. 6(a). When the electron concentration is small, $E^\prime_{R1}$ is a little lower than $E_{R1}$.

For generating a net spin-polarized photocurrent, both circularly-polarized light and the Rashba split bands are necessary\cite{ganichev2001conversion}. Upon illumination with a circularly polarized light with photon energy $\hbar\omega$ and given helicity, the energy and angular momentum conservations require that  the transition happens only at the two asymmetrical k values: $k^{+}_{x}$ and $k^{-}_{x}$\cite{ganichev2001conversion}. This makes the average electron velocity in the excited state become nonzero and the contributions of $k^{\pm}_{x}$ photoelectrons to the current do not cancel each other\cite{ganichev2001conversion}. Changing the photon helicity from +1 to -1 inverts the current because the "center-of-mass" for this transition is shifted in the opposite direction.
This results in the generation of the spin polarized CPGE current of the Rashba split d$_{yz}$/d$_{xz}$ bands, as shown in Figs. 6(b,c). The photon energies needed for the CPGE from the valence band edge ($E_t$) and the conduction band edge ($E^\prime_{R1}$) are shown in Fig. 6(d). In principle\cite{ganichev2001conversion,zhang2015generation,yuan2014generation}, it can be described by $j = \gamma \cdot$ \^{e}$E^{2}P_{circ}$, where $\gamma$ is the second-rank pseudotensor, $E$ is the amplitude of the electric field of the light, \^{e} is the unit vector pointing in the direction of the light propagation and $P_{circ}$ is the helicity of the light beam, and for the $C_{4v}$ point group, the  pseudotensor $\gamma$ has non-zero element, which results in a non-zero CPGE current\cite{moayed2017towards,vajna2012higher,golub2017photocurrent}.

\section{Conclusion}

In summary, through the first-principles calculations, we have systematically investigated the effect of the biaxial stress on the Rashba spin splitting of the KTO slabs for modelling strained KTO ultrathin films. The calculated results reveal that the Rashba spin splitting energy $E_R$ increases with the compressive stress increasing, which is in reasonable agreement with the recently experimental measurement, and $E_R$ becomes giant when the compressive in-plane strain  approaches $\epsilon_s=-5$\%. The largest $E_R$ is 140 meV and the corresponding k vector offset is $k_R=0.21$ {\AA}$^{-1}$, which implies that Rashba coupling constant is approximately 1.3 eV {\AA}. Compared to other systems, this $E_R$ is the next largest, only smaller than that in $\alpha$-GeTe(111) film \cite{krempasky2016disentangling}.
In contrast, the Rashba splitting changes little under tensile in-plane strain.
To elucidate the mechanism, we investigate the strain-dependent structural parameters and electrostatic potentials in the strained KTO slab.
For the unstrained KTO slab, there is a strong intrinsic electric field $e_0$ due to the out-of-plane alternate alignment of negative KO and positive TaO$_2$ monolayers. When compressive biaxial stress is applied, there is out-of-plane displacements of cations with respect to the neighboring anions driven by the compressive in-plane strain $\epsilon_s$ and tensile out-of-plane strain, and in addition the ionic displacements cause an out-of-plane electric field $e_d$ antiparallel to $e_0$. Our calculated results show that $e_d$ increases with  $\epsilon_s$, reaching the maximum nearly at $\epsilon_s=-5$\%.
Consequently, we can attribute the enhanced  giant Rashba spin splitting energy to the strong intrinsic out-of-plane electric field $e_d$ due to the large compressive biaxial stress.
Furthermore, our calculations show that the Rashba spin splitting can remain giant in the presence of STO capping and/or Si substrate. We also show that in addition to interesting quantum spintronic transports, such giant Rashba effect can be used to generate spin-polarized photocurrents in terms of the circular photogalvanic effect. Therefore, these giant Rashba phenomena may open a new door to promising spintronic and optoelectronic applications based on oxide thin films and heterostructures.


\begin{acknowledgments}
This work is supported by the Nature Science Foundation of China (Nos.11574366 and 11974393), by the Strategic Priority Research Program of the Chinese Academy of Sciences (Grant No.XDB07000000), and by the Department of Science and Technology of China (Grant No.2016YFA0300701). The calculations were performed in the Milky Way \#2 supercomputer system at the National Supercomputer Center of Guangzhou, Guangzhou, China.
\end{acknowledgments}


%

\end{document}